\documentclass[conference]{IEEEtran}

\usepackage[table,xcdraw]{xcolor}
\usepackage{float}
\usepackage{subfig}
\usepackage{amsmath,epsfig,psfrag} 
\usepackage{amsfonts,color,graphics,graphicx}
\usepackage{algorithm} 
\usepackage{algpseudocode}
\usepackage{multirow}
\usepackage{rotating}
\usepackage[numbers]{natbib}
\usepackage[numbers]{natbib}
\usepackage[normalem]{ulem}
\renewcommand{\baselinestretch}{1}

\usepackage{verbatim}

\begin{document}

\title{FRVM: Flexible Random Virtual IP Multiplexing in Software-Defined Networks}
	
\author{\IEEEauthorblockN{Dilli P. Sharma, Dong Seong Kim}
	\IEEEauthorblockA{
      University of Canterbury\\
		Christchurch, New Zealand \\
		dilli.sharma@pg.canterbury.ac.nz\\
        dongseong.kim@canterbury.ac.nz}
	\and
	\IEEEauthorblockN{Seunghyun Yoon, Hyuk Lim}
	\IEEEauthorblockA{
Gwangju Institute of Science \\ and Technology\\
		Gwangju, Republic of Korea\\
		\{seunghyunyoon, hlim\}@gist.ac.kr}
	\and
\centering
	\IEEEauthorblockN{ Jin-Hee Cho, Terrence J. Moore}
	\IEEEauthorblockA{
		Army Research Laboratory\\
		Adelphi MD, USA \\
		\{jin-hee.cho, terrence.j.moore\}.civ@mail.mil}
}
\maketitle             
	
\begin{abstract}
Network address shuffling is one of moving target defense (MTD) techniques that can invalidate the address information attackers have collected based on the current network IP configuration. We propose a software-defined networking-based MTD technique called {\em Flexible Random Virtual IP Multiplexing}, namely FRVM, which aims to defend against network reconnaissance and scanning attacks. FRVM enables a host machine to have multiple, random, time-varying virtual IP addresses, which are multiplexed to a real IP address of the host. Multiplexing or de-multiplexing event dynamically remaps all the virtual network addresses of the hosts. Therefore, at the end of a multiplexing event, FRVM aims to make the attackers lose any knowledge gained through the reconnaissance and to disturb their scanning strategy. In this work, we analyze and evaluate our proposed FRVM in terms of the attack success probability under scanning attacks and target host discovery attacks. \end{abstract} 	

\begin{IEEEkeywords}
	Network address shuffling, IP multiplexing, moving target defense, scanning attacks, attack success probability, software-defined networks
\end{IEEEkeywords}

\section{Introduction}
Conventional networked systems have been characterized by static system configurations which can greatly provide benefits to attackers in terms of their resource utilization in time and effort. The attackers often enjoy the asymmetric advantages because they can take enough time to investigate a target system by collecting its configuration information to identify exploitable vulnerabilities. Based on the obtained intelligence towards the system configuration, the attackers can plan to launch their attacks in order to maximize their utility and success.

The concept of moving target defense (MTD) has been introduced to increase uncertainty and/or confusion for attackers by continuously and dynamically changing the attack surface on a system. The dynamic defense based on MTD techniques has been often applied at a system level and/or a network level~\cite{Okhravi:MovingTT2016}. At the system level, the example MTD techniques aim to change system configurations in terms of instruction sets~\cite{Barrantes:RISEmulation2005, Kc:ISR2003, Portokalidis:GlobalISR2011}, run-time configuration~\cite{Xu:TransRuntime2003}, or IP address shuffling~\cite{Shacham:ASR2004}. At the network level, the common MTD technique is related to the shuffling of network attributes, including shuffling of MAC address, IP address, and/or Port number~\cite{Al-Shaer:RHMforMTD2013, Antonatos:NASRHitlistWorm2005, Atighetchi:APOD2003, Duan:RRM2013, Jafarian:AddMutation2015, Jafarian:AdversaryawareIP2015, Jafarian:OFRHM2012, Jafarian:SpatioMutation2014,  Jia:MOTAG2013, Kewley:DyNAT2001, Luo:RPAH2015, Luo:KeyedHash2017, MacFarland:SDNSuffle2015, Shi:PortAddressHopp2007, Yackoski:SDNA2011,Skowyra:MTD16,Wang:MTD17}. One of well-known network address obfuscation techniques is randomly and dynamically changing an IP address of a host~\cite{Al-Shaer:RHMforMTD2013, Jafarian:OFRHM2012, Jafarian:SpatioMutation2014}. The $1$-to-$1$ mapping technique from a real IP ({\em rIP}) to a virtual IP ({\em vIP}) is addressed in OpenFlow Random Host Mutation (OF-RHM)~\cite{Jafarian:OFRHM2012}, RHM~\cite{Al-Shaer:RHMforMTD2013}, and spatio-temporal address mutation~\cite{Jafarian:SpatioMutation2014}. One-to-one address mapping requires more {\em vIPs} to satisfy the mutation rate constraint and unpredictability for hosts in the network, which often result in lack of scalability due to a limited address space. 
	
In this paper, we propose an MTD technique called ~\textit{Flexible Random Virtual IP Multiplexing} (FRVM) in a software defined networking (SDN) environment which enables a host to have multiple, random, and/or time-varying virtual IP addresses ({\em vIPs}). The {\em vIPs} are changed randomly and dynamically in order to invalidate a target system's information collected by adversaries. FRVM creates short-lived {\em vIPs} to hide the real IP addresses ({\em rIP}s) of end-hosts. It allows a full range of {\em vIPs} to a host for different services by mapping in an $M-to-N$ manner where $M$ refers to the number of {\em rIP} and $N$ is the number of {\em vIP}. The {\em vIPs} are changed randomly and continuously where they can lower down predictability of the changing patterns of IP addresses by attackers. The {\em rIP} of an end-host remains unchanged and is transparent to the end user. This SDN-based MTD solution defends against network reconnaissance and scanning attacks. We consider scanning attacks used for gathering system configuration information before the actual attacks have launched. An attacker often uses a customized set of software tools to scan the target system to identify information such as operating system types, IP addresses, port numbers, running services, protocols, network topology, and/or exploitable vulnerabilities. The proposed FRVM provides high network diversity in terms of IP addresses of end-hosts that can significantly disturbs the attacker's scanning strategies and accordingly reduces the attacker's success rate. 
This work makes the following {\bf key contributions}:
\begin{itemize} 
	\item We presented an efficient novel SDN-based MTD mechanism that provides flexibility to have multiple, random, time-variant IP addresses in a host, and this creates an environment with high diversity that significantly decrease the attacker's scanning success rate;
	\item We formulated the architectural framework and outline communication protocols of FRVM, in which it is feasible to  be implemented on an SDN environment; and
	\item We derived probabilistic models for analyzing and evaluating the effectiveness of FRVM; and our experimental results clearly support the outperformance of our proposed FRVM compared to the baseline and existing counterpart, with respect to defense strength. 
\end{itemize}
The rest of this paper is structured as follows. Section \ref{sec:related_work} provides the overview of SDN environments and IP shuffling techniques as an MTD mechanism. Section \ref{sec:system_model} discusses the network and threat models considered in this work. Section \ref{sec:frvm} describes the overall design of the proposed FRVM in terms of IP mapping, architecture, and communication protocol. Section \ref{sec:probability_model} provides a probability model for the analytical validation of the proposed FRVM. Section \ref{sec:results_analysis} demonstrates the experimental results and discusses their overall trends with physical interpretations. Lastly, Section \ref{sec:conclusion} concludes this work and suggests the future work directions.
	
\section{Background \& Related Work} \label{sec:related_work}

\subsection{Software-Defined Networking} \label{sec:sdn_background}
A conventional network consists of heterogeneous components, such as switches, firewalls, and routers with their own proprietary software and protocols. This kind of network setup is rarely flexible, which significantly hurdles applying diverse, novel ideas to optimize system performance and security. Software-defined networking (SDN) has emerged to mitigate the issues with which the existing traditional networks have faced~\cite{Sezer:AreWeReadySDN2013} by decoupling the network control and the forwarding functions that provide high programmability and efficient abstraction of applications and network services in the underlying infrastructure~\cite{Kreutz:SDNSurvey2015}. An SDN environment consists of SDN switches and controllers, communicating over a secure channel. Many SDN switch protocols have been developed in which one of well-known protocols is an {\em OpenFlow}~\cite{McKeown:2008OpenFlow} protocol, a fundamental element for building the SDN environment. The SDN's programmable interfaces help build dynamic, proactive, and adaptive defense mechanisms such as MTD techniques.

\subsection{IP Shuffling} \label{subsec:ip_shuffling_related_work}
Network address shuffling is one of MTD techniques whose key approach is to dynamically and frequently change IP addresses and port numbers of a target system. \citet{Antonatos:NASRHitlistWorm2005} proposed a network address space randomization scheme (NASR) to deal with hitlist worms. NASR is a local area network (LAN)-level network address randomization technique based on the dynamic host configuration protocol (DHCP) updates. Although NASR reflects the nature of shuffling based on the IP address randomization as an MTD mechanism, it disrupts the active connections and has limitations in providing high unpredictability and mutation speed due to constrained address space in terms of a LAN address. \citet{Shi:PortAddressHopp2007} developed an active cyber-defense mechanism based on port and address hopping with timestamp-based synchronization. However, this cyber-defense mechanism is developed to deal with Denial-of-Service (DoS) attacks only within the scope of metropolitan area networks (MANs) and multiple LANs.

Various types of IP address shuffling techniques have been proposed in the literature. Open-Flow Random Host Mutation (OF-RHM)~\cite{Jafarian:OFRHM2012} is an IP address shuffling technique that randomly mutates IP address of the end-hosts. In OF-RHM, the real IP address of the end-hosts remain unchanged while associating with a short-lived virtual IP address ({\em vIP}) where {\em vIPs} are randomly and periodically changed. The translation of {\em vIP}-to-{\em rIP} is performed right before the end-host. A {\em vIP} is chosen from the range after each mutation interval. The {\em vIP} is uniformly selected at random or with an associated weight. The OF-RHM is implemented on an SDN environment, in which each OF-switch performs {\em vIP}-to-{\em rIP} and {\em rIP}-to-{\em vIP} translations as specified by an SDN controller. Although the OF-RHM is transparent to end-hosts and provides the high mutation rate, it cannot be deployed in a conventional network due to low scalability.

\citet{Al-Shaer:RHMforMTD2013} proposed Random Host Mutation (RHM) to solve the scalability problem in conventional networks. RHM uses a two-phase mutation approach consisting of Low Frequency Mutation (LFM) and High Frequency Mutation (HFM) intervals to assign {\em vIP}. LFM is used to optimally select a random network address for moving target hosts (i.e., denoted by virtual address range, or VAR) under constraints while HFM is used to select random {\em vIP} within VAR assigned during LFM. The virtual address space allocation is performed by the moving target controller and the translation between {\em rIP} and {\em vIP} is performed by moving target gateway (MTG). The RHM is implemented in a traditional campus network with the validation of its effectiveness in dealing with internal and external scanners. However, this work didn't consider the effectiveness of the RHM against more intelligent attacks such as Advanced Persistent Threats (APTs).

\citet{Jafarian:SpatioMutation2014} also presented a spatio-temporal address mutation technique which aims to mitigate the impact of sophisticated APTs in enterprise networks. This scheme presented dynamicity into a network by invalidating the information attackers collected during the reconnaissance at one host that can be used at other time and other hosts by varying host IP binding dynamically based on location and time. Each host is associated with the unique set of IP address called `ephemeral IP' ({\em eIP}) to reach other hosts in the network. This approach is implemented on a legacy network; but its effectiveness is evaluated on the SDN environment. The translation of {\em rIP}-to/from-{\em eIP} is performed in the gateway of the conventional network, and OF-Switch of the SDN environment. This work investigated the effectiveness of their scheme using three metrics, including deterrence, deception, and detectability, under attacks such as cooperative and local preference worms and/or APT attacks. 
	
In this work, we propose the FRVM as an MTD mechanism in order to map IP addresses in an $M$-to-$N$ manner, which provides high flexibility in IP mapping while generating high diversity to enhance security. We validate the outperformance of FRVM in terms of reducing attack success rate under scanning attacks and accordingly discovering target hosts. Our experimental results show that the proposed FRVM significantly provides an effective MTD mechanism by providing high network diversity, ultimately leading to enhancing system and network security.

\begin{figure*}[ht!]
	\centering	\includegraphics[width=1.0\textwidth]{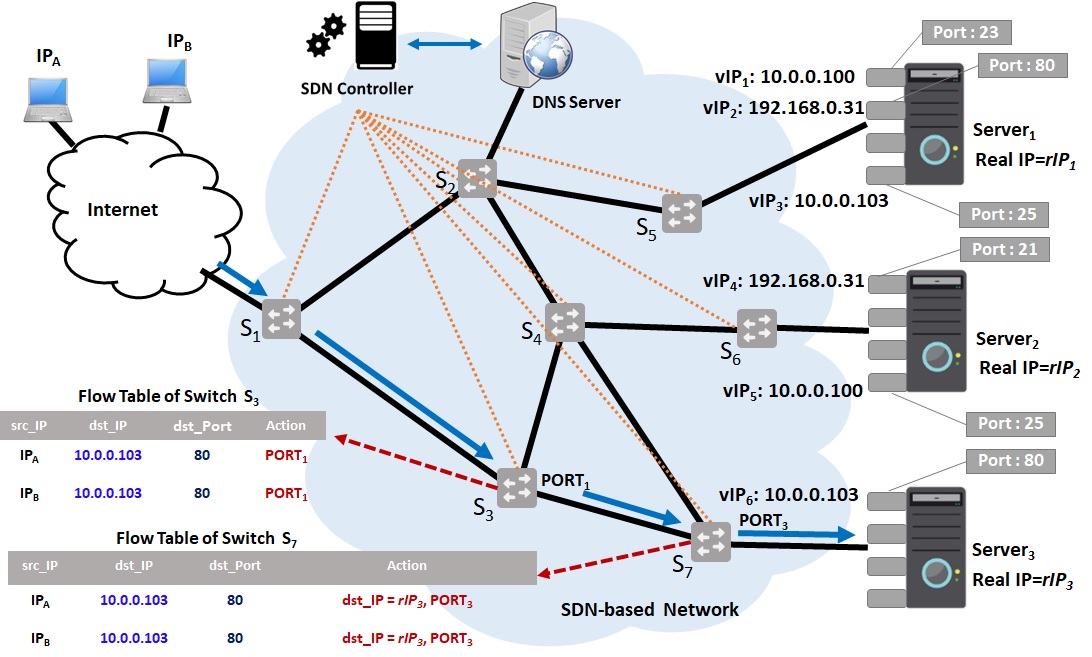}
	\caption{FRVM implementation on an SDN-based environment.}
	\label{fig_systemarchitecture}
\end{figure*} 
	
\section{System Model} \label{sec:system_model}
In this section, we discuss our network and threat models that describe assumptions made and attack behaviors considered in this work.

\subsection{Network Model} \label{subsec:network_model}	
We consider an SDN-based environment in which end-hosts are called server-hosts which offer a number of services to end-users. Each server-host should have at least one active service where a server can have multiple services. The active services running on server-hosts are identified by the port numbers, and a {\em vIP} is assigned to each service. A combination of {\em vIP} and port number (i.e., {\em vIP: Port}) can be used to communicate within the SDN-based network environment. Fig.~\ref{fig_systemarchitecture} describes our system model consisting of the control plane with a controller (e.g., an SDN controller) and the data plan with switches and hosts (e.g., OpenFlow switches and server-hosts). 
	
\subsection{Threat Model} \label{subsec:threat_model}
An attacker seeks to perform reconnaissance on the network while the system defense mechanism aims to hide the IP addresses and port numbers. The attacker selects a random IP address and port number to discover the host (i.e., IP address) and active services (i.e., ports) on the host for planning and launching attacks. In this work, an attacker is assumed to have some level of knowledge towards a target system and its defense capability as follows: 
\begin{itemize}
\item $n$ number of hosts (i.e., the number of servers) exist where each host has a real IP address, {\em rIP}.
\item $m$-active services are running on each server host. 
\item An attacker targets a service running on a server-host by performing the target server's {\em vIP} and port scanning. 
\item The attacker is aware of the virtual address space pool and will sequentially attempt $k$ connections or probes.
\item The attacker uses a scanner (e.g., $Nmap$~\cite{Lyon:Nmap2009}) to discover active server hosts and active services running on the target end-host.  
\item The multiplexing event dynamically remaps all {\em vIPs} at the server host by uniformly selecting {\em vIPs} at random.
\end{itemize}	

\section{Flexible Random Virtual IP Multiplexing} \label{sec:frvm}
In this section, we present our proposed MTD mechanism, called \underline{F}lexible \underline{R}andom \underline{V}irtual IP \underline{M}ultiplexing, namely FRVM, that provides the network address diversity by multiplexing and de-multiplexing IP addresses of a target end-host. 
	  
\subsection{IP Mapping} \label{subsec:ip_mapping}
In FRVM, each server-host has a real IP address, {\em rIP}, where the {\em rIP} is mapped to a set of virtual IP addresses, {\em vIPs}, and a set of {\em vIPs} are mapped to an {\em rIP} in an $1$-to-$n$/$n$-to-$1$ manner, respectively.  The mapping of one or more {\em vIPs} to an {\em rIP} (i.e., {\em vIPs}-to-{\em rIP}) of each server-host is called `virtual IP multiplexing' while the mapping of a {\em rIP} to {\em vIPs} mapping is called `de-multiplexing.' Real IP addresses are public IP addresses while the {\em vIPs} are unused, private IP addresses (e.g., 10.0.0.0/24, 172.16.0.0/20 or 192.168.0.0/16). We define the multiplexing function in Eq.~\eqref{equation:vIP-Multiplexing} that describes the multiplexing of a set of {\em vIPs} (e.g.,$vIP_{1}, vIP_{2}, \cdots, vIP_{m}$) to a set of {\em rIPs} (e.g., a $rIP_{1}$) in $m$-to-$1$ manner. Similarly, the de-multiplexing function in Eq.~\eqref{eq:frvmmap} details the de-multiplexing of a set of {\em rIPs} (e.g., $rIP_{1}$) to a set of {\em vIPs} (e.g., $vIP_{1}, vIP_{2}, \cdots, vIP_{n}$) in an $1$-to-$n$ manner. To perform multiplexing/de-multiplexing of IP addresses,  Eqs.~\eqref{equation:vIP-Multiplexing} and \eqref{eq:frvmmap} combine a {\em vIP} with a port number (e.g., $vIP: port$).
\begin{equation} \label{equation:vIP-Multiplexing}
f_{multiplexing}: \mathbb{VIP} \rightarrow \mathbb{RIP}
\end{equation}  
\begin{equation} \label{eq:frvmmap}
f_{de-multiplexing}: \mathbb{RIP} \rightarrow \mathbb{VIP}
\end{equation} 
As described in Section~\ref{subsec:threat_model}, each server-host offers one or more services, and these services are identified with the corresponding port number. A new {\em vIP} is generated randomly using a cryptographically secured random number generator and assigned to the server-host's service. All the randomly assigned {\em vIPs} to the server-host are changed continuously based on a short period of the interval, called `multiplexing interval,' denoted by $T$. The {\em rIP} of the server-host remains unchanged while {\em rIP} is hidden in the network and routing is restricted to using {\em vIPs}. The SDN controller handles all the mapping and remapping of addresses, changing of {\em vIPs} of the server-host and updating the flow-table entry of SDN switches (e.g., Open-Flow-Switches). The controller performs mapping of {\em vIPs}-to-{\em rIP} or {\em rIP}-to-{\em vIP} at edge switch (i.e., just before the end-host). The mapping is transparent to an end user with no service disruption since {\em rIPs} and corresponding port numbers of a server-host remain unchanged. We assume that no one can reach to the server-host using the {\em rIP} except the authorized network administrators. A user can communicate either using a domain name/host-name or {\em rIP}, which are described as the communication protocol in FRVM in Section~\ref{subsection:protocol}.
  
\subsection{System Architecture} \label{subsec:system_architecture}
FRVM is implemented on an SDN-based environment, as described in Fig.~\ref{fig_systemarchitecture}. The FRVM architecture consists of the following components: end-hosts, an SDN controller, SDN switches, data plane, and control plane. The end-hosts are connected with the SDN-switches in the data plane. Data forwarding network elements are centrally controlled by the SDN controller. The components of the FRVM system architecture can be represented by:
\begin{itemize}
\item \textit{End-host(s)}: A set of end-hosts includes servers, workstations, and/or gateway interface to routers, denoted by~$\mathbb{H}=\{Host_{1},Host_{2}, \cdots, Host_{n}\}$ where $|\mathbb{H}| \geq 2$. Each end-host has an {\em rIP} assigned to it and a set of  {\em rIPs} is denoted by~$\mathbb{RIP} =\{rIP_{1}, rIP_{2}, \cdots, rIP_{n}\}$. A {\em rIP} is mapped to one or more {\em vIPs}, and a set of all {\em vIPs} is denoted by~$\mathbb{VIP} = \{vIP_{1}, vIP_{2}, \cdots, vIP_{m}\}$. End-host offers a number of active services listening on port numbers.
\item \textit{SDN Controller}: A set controllers, denoted by~$\mathbb{C}=\{C_{1}, C_{2}, \cdots, C_{m}\}$. We assume that a functional SDN network has at least one controller where $|\mathbb{C}|\geq 1$. In our system, a single SDN controller exists to perform the mapping of {\em rIP}-to-{\em vIP} or {\em vIP}-to-{\em rIP}, and install necessary flows in the SDN-switches. The SDN controller randomly and dynamically changes {\em vIPs} of end-hosts and {\em rIPs} of the end-host machines remain unchanged. 
\item \textit{SDN-switches}: A set of SDN-witches, denoted by~$\mathbb{S} =\{S_{1},S_{2}, \cdots, S_{k}\}$. Each SDN-switch forwards the data plane traffics based on the flow-rule specified by the SDN controllers. Each switch,~$S_{i}$, includes ingress and egress ports.
\item \textit{Data plane}: Representing a topological connectivity based on a set of network switches and end-hosts.
\item \textit{Control plane}: Representing relations between switches and controllers. 
\item \textit{End-users}: A set of legitimate users, denoted by~$\mathbb{U}$=$\{u_{1},u_{2}, \cdots, u_{j}\}$ that access the services running on the servers in~$\mathbb{H}$.
\item \textit{DNS servers}: The domain name system (DNS) server resolves a domain name of the server-host and returns a {\em rIP} address. 
\end{itemize}

The IP multiplexing algorithm for an SDN-controller is presented in Algorithm~\ref{algo:ipmultiplexing}. The SDN-switch sends every unmatched packet to the controller which determines the types of communication (e.g., domain-name or {\em rIP}), performs the translation of IP addresses, and installs necessary flows. The SDN-controller maps a real IP address of the destination host to a number of virtual IP addresses and then updates the flow table of each SDN-switch (e.g., OF-switch). For example, Table~\ref{tbl_flow-table_s1} shows the state of a flow-table of an OF-switch where destination {\em rIPs} are replaced with the {\em vIPs}. Table~\ref{tbl_address_multiplexing} shows the mapping of $1$-to-$n$/$n$-to-$1$ IP addresses using this algorithm. All {\em rIPs} hidden within the SDN network packets are forwarded using the {\em vIPs} only. The SDN controller dynamically changes {\em vIPs} for each service of a server-host in every multiplexing time interval, $T$. 
\begin{algorithm}[ht!]
	\caption{Algorithm for Multiplexing IP Addresses}
	\label{algo:ipmultiplexing}
	\begin{algorithmic}[1]
		\For {packets $p$ from host {$h_{i}$ to $h_{j}$}}
			\If {$p$ is a Type-A DNS response for host $h_{j}$}
				\State Set \textit{dstAddr(p):=vIP($h_{j}$):dstPort}
			\Else 
			\If {$h_{i}$ is authorized access to $h_{j}$}
				\State Set \textit{dstAddr(p):=vIP($h_{j}$):dstPort}
			\EndIf
			\EndIf
			\If {$p$ is at source SDN-Switch}
				\State Set \textit{scrAddr(p):=vIP($h_{i}$):srcPort}
				\State Set \textit{dstAddr(p):=rIP($h_{i}$):dstPort}
			\EndIf
			\If {$p$ is at destination SDN-Switch}
				\State Set \textit{dstAddr(p):=rIP($h_{j}$):dstPort}
				\State Set \textit{scrAddr(p):=vIP($h_{j}$):srcPort}
			\EndIf
		   \For {each service of a host $h_{i}$}
		   		\State Select a new {\em vIP} for each service of $h_{i}$
		   		\State Update flow table entry
		   \EndFor
		 \EndFor
	\end{algorithmic}
\end{algorithm}

\begin{table*}[th]
	\centering
	\caption{Flow-table for a OF-Switch }
	\label{tbl_flow-table_s1}
	\begin{tabular}{|c|c|c|c|}
		\hline
		\rowcolor[HTML]{C0C0C0} 
		\textbf{Src\_IP} & \textbf{Dst\_IP} & \textbf{Dst\_Port\#} & \textbf{Action}                                              \\ \hline
		$IP{_A}$   & 10.0.0.103      & 80               & Forward to $PORT{_1}$ with Dst\_IP  10.0.0.103 and Dst\_Port\#80 \\ \hline
		$IP{_B}$   & 10.0.0.103      & 80               & Forward to $PORT{_1}$ with Dst\_IP 10.0.0.103 and Dst\_Port\#80  \\ \hline
		$IP{_C}$   & 10.0.0.100     & 23                   & Forward to $PORT{_2}$ with Dst\_IP 10.0.0.100 and Dst\_Port\#23  \\ \hline
		$IP{_D}$   & 192.168.0.31      & 21                  & Forward to $PORT{_3}$ with Dst\_IP 192.168.0.31 and Dst\_Port\#21  \\ \hline
	\end{tabular}
\end{table*}

\begin{table}[th!]
	\centering
	\caption{Example of IP address De/multiplexing}
	\label{tbl_address_multiplexing}
	\begin{tabular}{lll} 
		\rowcolor[HTML]{C0C0C0} 
		\hline
		\multicolumn{1}{|c|}{\cellcolor[HTML]{C0C0C0}Real\_IP} & \multicolumn{1}{c|}{\cellcolor[HTML]{C0C0C0}TCP\_Port\#} & \multicolumn{1}{c|}{\cellcolor[HTML]{C0C0C0}Virtual\_IP} \\ \hline
		\rowcolor[HTML]{FFFFFF} 
		\multicolumn{1}{|c|}{\cellcolor[HTML]{FFFFFF}} & \multicolumn{1}{c|}{\cellcolor[HTML]{FFFFFF}23} & \multicolumn{1}{c|}{\cellcolor[HTML]{FFFFFF}10.0.0.100} \\ \cline{2-3} 
		\rowcolor[HTML]{FFFFFF} 
		\multicolumn{1}{|c|}{\cellcolor[HTML]{FFFFFF}} & \multicolumn{1}{c|}{\cellcolor[HTML]{FFFFFF}80} & \multicolumn{1}{c|}{\cellcolor[HTML]{FFFFFF}192.168.0.31} \\ \cline{2-3} 
		\rowcolor[HTML]{FFFFFF} 
		\multicolumn{1}{|c|}{\multirow{-3}{*}{\cellcolor[HTML]{FFFFFF}$rIP{_1}$}} & \multicolumn{1}{c|}{\cellcolor[HTML]{FFFFFF}25} & \multicolumn{1}{c|}{\cellcolor[HTML]{FFFFFF}10.0.0.103} \\ \hline
		\rowcolor[HTML]{FFFFFF}
		\multicolumn{1}{|c|}{\cellcolor[HTML]{FFFFFF}} & \multicolumn{1}{c|}{\cellcolor[HTML]{FFFFFF}21} & \multicolumn{1}{c|}{\cellcolor[HTML]{FFFFFF}192.168.0.31} \\ \cline{2-3} 
		\rowcolor[HTML]{FFFFFF}
		\multicolumn{1}{|c|}{\cellcolor[HTML]{FFFFFF}$rIP{_2}$} & \multicolumn{1}{c|}{\cellcolor[HTML]{FFFFFF}25} & \multicolumn{1}{c|}{\cellcolor[HTML]{FFFFFF}10.0.0.100} \\ \hline
		\rowcolor[HTML]{FFFFFF} 
		\multicolumn{1}{|c|}{\cellcolor[HTML]{FFFFFF}$rIP{_3}$} & \multicolumn{1}{c|}{\cellcolor[HTML]{FFFFFF}80} & \multicolumn{1}{c|}{\cellcolor[HTML]{FFFFFF}10.0.0.103} \\ \hline
		\end{tabular}
\end{table}

\subsection{Communication Protocol}
\label{subsection:protocol} 
\begin{figure*}[ht!]
\centering
		\includegraphics[width=0.96\textwidth]{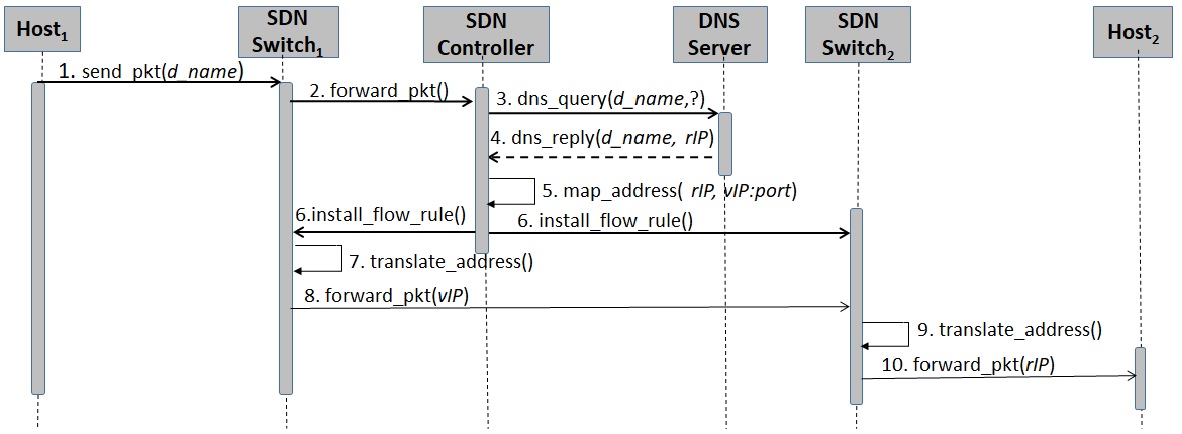}
		\caption{Communication via domain-name.}
		\label{fig_protocol_domain}
\end{figure*}

\begin{figure*}[ht!]
	\centering
	\includegraphics[width=0.96\textwidth]{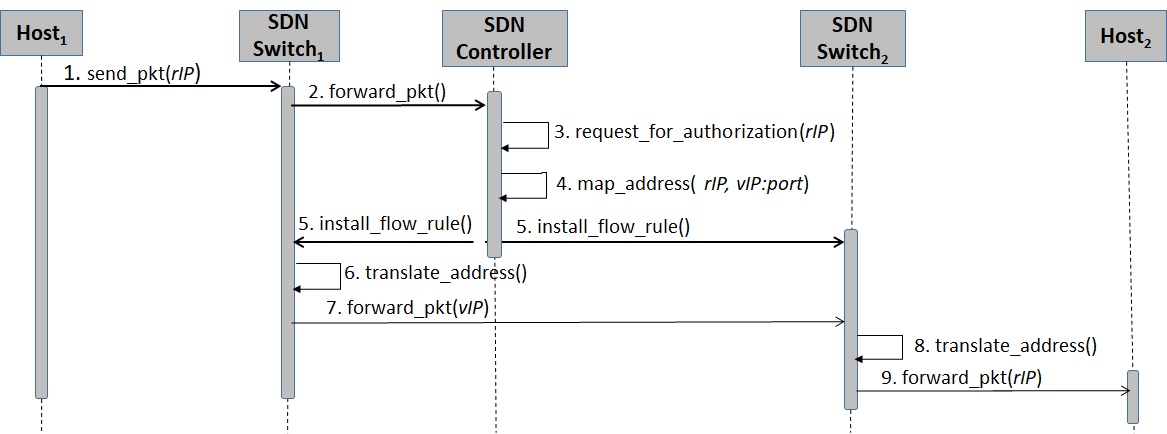} 
	\caption{Communications via real IP.}
	\label{fig_protocol_rip}
\end{figure*}

There are two ways of communicating either using 'domain-name' or {\em rIP}, which is shown in Figs.~\ref{fig_protocol_domain} and ~\ref{fig_protocol_rip}, respectively. Fig.~\ref{fig_protocol_domain} shows that when a DNS query is sent to resolve the domain-name (e.g, $d_{\_}name$) of an end-host (e.g., $Host_{2}$), then the DNS response is intercepted by the SDN controller and the {\em rIP} of the destination host (e.g, $rIP_{2}$) is replaced with the virtual IP address (e.g., $vIP_{1}$). As a result, the source end-host (e.g., $Host_{1}$) receives only a {\em vIP} (e.g., $vIP_{2}$) mapping to the destination host and initiates its communication with the destination service port number (e.g., $Port_{2}$). Similarly, a source host's {\em rIP} (e.g., $rIP_{1}$) of an outgoing packet is replaced by a randomly selected {\em vIP} (e.g., $vIP_{1}$). The SDN controller installs necessary flows in the SDN-switches in the route. Further, packets will be matched and forwarded by the SDN-switches according to the installed flows in the flow table. The translation of the {\em vIP}-to-{\em rIP} will be applied at the SDN-switch before the destination end-host. 

Fig.~\ref{fig_protocol_rip} shows how an authorized user at a host (e.g., $Host_{1}$) can reach an end-host (e,g., $Host_{2}$) using the real IP address. We assume that only an authorized user (e.g., a network administrator) is allowed to communicate using {\em rIP} of the destination host. In this scenario, the client initiates connection using {\em rIP} (e.g., $rIP_{2}$) and port number (e.g., $Port_{2}$) of the destination host (e.g., $Host_{2}$), and the SDN controller requesting authorization for a received packet from the source host (e.g., $Host_{1}$). If an access is granted, the SDN controller replaces the {\em rIP} of the destination host (e.g., $rIP_{2}$) with a randomly generated {\em vIP} (e.g., $vIP_{2}$), and then sends the packet to the SDN controller which installs necessary flows in the SDN-switches. The authorization is assumed to be performed per TCP or UDP connection. In the FRVM network, routing is restricted to use only {\em vIPs} while {\em rIPs} are hidden within the network. The multiplexing and de-multiplexing functions enable the association between the destination port and the IP address.  


\section{Probabilistic Derivation of Attack Success Probability}	\label{sec:probability_model} 
For the IP configuration in a static network, the IP addresses assigned to end-hosts remain unchanged in the network. An attacker's strategy for discovering a target host is to sequentially scan and iterate through the address space. The attack success probability then is determined via hyper-geometric distribution as the number of success in a sequence of $k$-draws from a finite population without replacement. Accordingly, the probability that the attacker successfully obtains exactly $x$ of $n$ hosts in the address space $N$ in $k$ scans can be given by:
\begin{equation} \label{eq:static-k-target-asp}
P(X=x)={\frac{{\binom{n}{x}}\binom{N-n}{k-x}}{\binom{N}{k}}} 
\end{equation}
where $k \geq x$, $N \geq n$, and $n \geq x$. The probability that the attacker successfully discovers at least one target host is:
\begin{equation} \label{eq:static-one-target-asp}
P(X>0)=1-P(X=0)=1-{\frac{\binom {N-n}{k}}{\binom {N}{k}}}
\end{equation}
In our work, an address shuffling event (e.g., multiplexing or de-multiplexing event) remaps all {\em vIPs} to the end-hosts. We assume that FRVM remaps all the IP addresses of hosts in every scanning attempt where the equal rate for shuffling and scanning is assumed. An attacker loses any reconnaissance knowledge gained in every shuffling event. The attacker success probability remains same upon every scanning. The attacker has to scan all the IP address in each scan for discovering the target end-host. Due to this reason, the attacker success probability (ASP) is determined through a binomial distribution as the number of success in a sequence of $k$-draws from a finite population with replacement. The binomial probability distribution function with the success probability $p$ is:
\begin{equation} \label{eq:binomial}
P(X=x)={\binom {k}{x}}p^x(1-p)^{k-x}
\end{equation}
where $p=\frac{n}{N}$ is the probability that the attacker discovers a host and $k \geq x$. This binomial probability function in Eq.~\eqref{eq:binomial} can be used to estimate the attacker success rate, given the network address space ($N$), the number of hosts ($n$), and the number of scans ($k$) in this work. Therefore, the ASP for an attacker to discover a host in $k$ scans is given by:
\begin{eqnarray} \label{eq:binomial-single-target-asc}
P(X>0) = 1-P(X=0) = 1-(1-p)^k = 1-(1-\frac{1}{N})^k \nonumber
\end{eqnarray} 
If the attacker scans whole address space (i.e., $k=N$), then Eq.~\eqref{eq:binomial-single-target-asc} reduces to
\begin{equation}
P(0<X\leq N)=1-(1-\frac{1}{N})^N.
\end{equation} 
As the network address space $N$ increases to a sufficiently large number (i.e., $N\rightarrow\infty$), the ASP converges to $1-(1-\frac{1}{N})^N =1-e^{-1} \approx 0.63$. 

Denote a multiplexing interval by $T$, a multiplexing rate of each server host by $\theta$ where $\theta=\frac{1}{T}$, a scanning rate by $\eta$ (i.e., the number of probing packets that a scanner host sent out per second), the ratio of attacker scanning rate over defender multiplexing rate by $r$ where $r=\frac{\eta}{\theta}$, and the number of scans by $k$. If the scanning rate is as same as the multiplexing rate (i.e., $r=1$, the multiplexing event reshuffles IPs per scan), the attacker will miss the target host with probability $(1-\frac{1}{N})$ for every scan and probability $(1-\frac{1}{N})^k$ in $k$-scans. If the attacker scans entire address $k=N$, then the attacker will miss the host with probability $(1-\frac{1}{N})^N =e^{-1} \approx 0.37$. If $r<1$, the attacker cannot scan the whole address space in each multiplexing interval, $T$. 
If the scanner rate is greater than the multiplexing rate with $r>1$ (i.e., $k>N$), the attacker will be able to scan all the addresses.

\section{Results \& Analysis}	\label{sec:results_analysis}

\subsection{Experimental Setup} \label{subsec:exp_setup} 
In this work, we focus on evaluating the effectiveness of FRVM under scanning attacks with respect to the attack success probability. As discussed in Section \ref{subsec:threat_model}, the attacker seeks to perform reconnaissance on the network while defense mechanism aims to keep the IP addresses (e.g., {\em vIPs}) hidden. An attacker selects a random IP address to discover a target server-host, and then performs port scanning on the host to discover the active services. We consider a random scanning strategy (e.g., non-repeat scanning) to discover IP addresses of the target server hosts and active service running ports of the target server host. The active server-host discovery (e.g., IP addresses) can be achieved through sending an ICMP message to the target server host. Similarly, discovering the active services can be achieved by sending a SYN packet to the target server service using port scanning. The attacker success probability (e.g., scanner success rate) and overhead (e.g., flow-table size) are used to evaluate the effectiveness of the FRVM and its counterpart, a static network that uses a static IP configuration.

The attacker success probability (ASP) is affected by various environmental parameters, such as the address space size, number of scans, scanning rate, and/or shuffling rate of defense mechanism. The ASP is obtained based on Eqs.~\eqref{eq:static-one-target-asp} and \eqref{eq:binomial-single-target-asc} for both the static network configuration and the network with FRVM, respectively. 
 
\subsection{Results} \label{section:result_analysis}
\begin{figure*}[!ht]
\centering
\subfloat[ASP with the static network and FRVM for a single target]{
\label{fig_asp_singletarget}
\includegraphics[width=0.5\textwidth, height=0.3\textwidth]{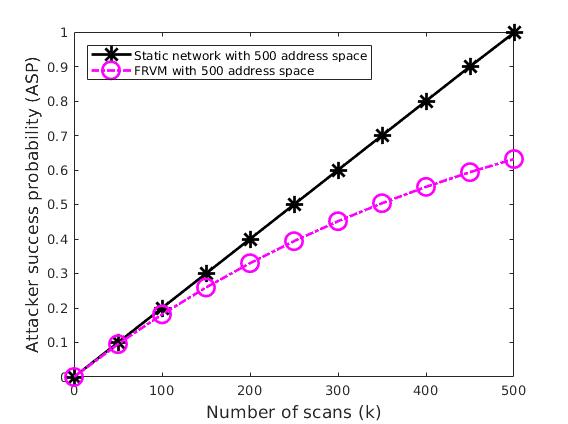}}
\subfloat[ASP with all {\em vIPs} of Class-A, Class-B and Class-C]{
\label{fig_asp_classesIP}
\includegraphics[width=0.5\textwidth, height=0.3\textwidth]{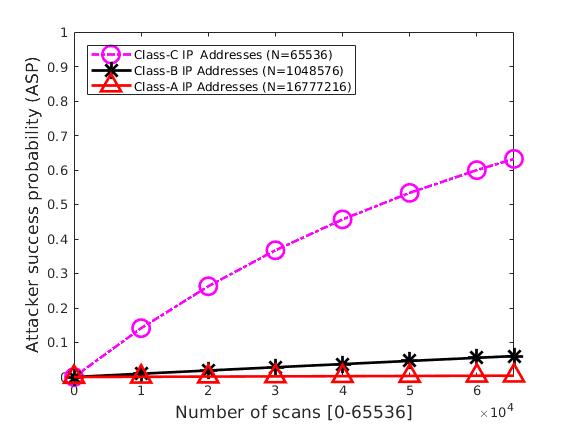}}

\subfloat[ASP with static network and FRVM for address space $2^{16}$]{
	\label{fig_asp_vIP_of_classC}
	\includegraphics[width=0.5\textwidth, height=0.3\textwidth]{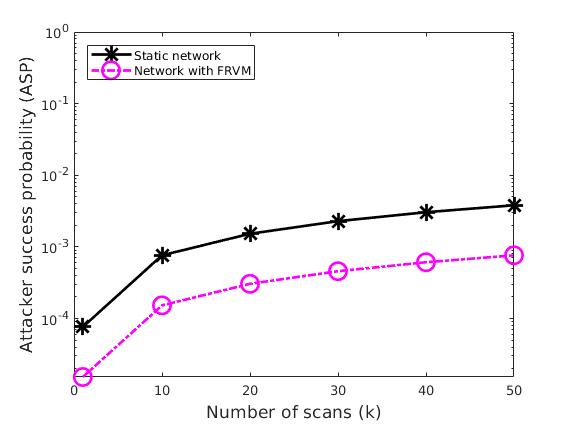}}
\subfloat[ASP with the static network and FRVM for multiple host dicovery]{
\label{fig_asp_multiple_discovery}
\includegraphics[width=0.5\textwidth, height=0.3\textwidth]{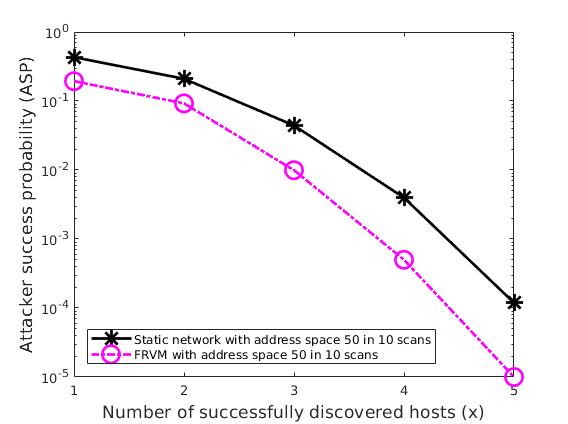}}

\caption{Comparative performance analysis with respect to the scanning attack success probability (ASP)}
\label{fig_asp}
\end{figure*}

{\bf Effect of scanning attacks}: Fig.~\ref{fig_asp} (a) shows that the ASP increases as the number of scans increases. For the static network, the ASP reaches to 1.0 (i.e., 100\%) when the attacker scanned the whole address space. However, in the network with FRVM, the ASP slightly increases as the number of scans increases; the ASP reaches up to 0.63 (i.e., 63\%) when the attacker scanned the whole address space. This implies that the attacker missed 37\% of the hosts when they scanned the whole network addresses. 

{\bf Effect of discovering a virtual IP address attack}: Similarly, Fig.~\ref{fig_asp} (b) depicts the ASP when the attackers aim to discover a virtual IP address in the range of IPv4 Class-A (i.e., 10.0.0.0/24), Class-B (i.e.,172.16.0.0/20 ), and Class-C (i.e., 192.168.0.0/16) with respect to the number of scans. Fig.~\ref{fig_asp} (b) shows that the scanning success probability is 0.63 in scanning all the virtual IP addresses of Class-C (i.e., $2^{16}=65,536$). A large number of scans ($k$) is required to discover the virtual IP addresses of Class-B; furthermore, extensively a large number of scans is required to discover the virtual IP address of Class-A while the attacker missed 37\% IP addresses even when scanning all the IP addresses (i.e., $2^{16}+2^{20}+2^{24}=17,891,328$).

{\bf ASP under a single target host}: Let us consider another example with five server-hosts ($n=5$) and initial five IP addresses where $vIP_{1}, vIP_{2}, vIP_{3}, vIP_{4}$ and $vIP_{5}$ are randomly assigned to each host, $Host_{1}, Host_{2}, Host_{3}, Host_{4}$, and $Host_{5}$ from address space ($N=65,536$, 192.168.00.00/16), respectively. The ASP computed for both networks, the static network with no MTD and the network with FRVM, given the number of scans ($k$). The result in Fig.~\ref{fig_asp} (c) shows that the network with FRVM reduces ASP five times on average (i.e., $\approx$ 4.99 times), compared to ASP observed in the static network.

{\bf ASP to discover multiple hosts}: Fig.~\ref{fig_asp} (d) shows ASP to discover the number of hosts in a network. In this experiment, we setup a small network with 5 server hosts ($n=5$) and 50 address space ($N=500$) and computed ASP to discover 1 host, 2 hosts, 3 hosts, 4 hosts and finally all 5 hosts for both the static network and the network with FRVM in 10 scans ($k=10$). The result shows that FRVM gradually reduces ASP as compared to the static network.

\subsection{Discussion} \label{subsec:discussion}	

The results in Section~\ref{sec:results_analysis} show that FRVM can effectively thwart scanning attacks by invalidating at least 37\% of the hosts/services in the worse-case scenario (i.e., scanned whole address). These results are obtained for an equal address space pool used by both of the attacker and defender to scanning and multiplexing IP addresses, respectively. However, FRVM updates only the small portion ($q$) of the address space during a scanning time of the attacker. Therefore, the multiplexing rate of FRVM is higher than the attacker scanning rate. In this scenario, ASP in terms of discovering a target host decreases and becomes very low. For example, if the scanner will be able to scan only 10\% of the address space ($N$) and FRVM multiplexes all the IP addresses in this time, ASP will be 9.5\% (i.e., $ASP=0.095$). This implies that FRVM can effectively protect at least 90.5 \% of the hosts/services from discovery. Similarly, if the scanner scans only 5\%  of the total address space, ASP reduces to 4.8\% (i.e., $ASP=0.048$); and for the 1\% scan, ASP further reduces to 0.9\% ($\approx 1\%$) (i.e., $ASP=0.0099$). This all shows that FRVM effectively thwarts about 99\% network reconnaissance of the hosts/services. 

FRVM also increases the address space diversity by adding the flexibility to have multiple {\em vIPs} per services on a host in the network. The environment with high address space diversity increases confusion or uncertainty for the attacker to identify a designated target host. 

In a static SDN-based network, an SDN controller makes an entry for each host in the flow-table of OF-switch. The size of the flow table increases as the number of hosts increases. In FRVM, the number of IP addresses assigned to a host is dependent upon the running serves. $m$ denotes an average number of active services of each host at a particular point of time; and $n$ denotes the number of end-hosts in the SDN-based network. Therefore, the number of flow-table entries in OF-switch is $n \times m$, and the size of the flow-table changes as the flow-table is being updated. FRVM adds some operational delay due to the mapping and remapping IP addresses. 

For multiplexing and de-multiplexing, virtual and real IP addresses should be converted to the real and virtual IP addresses, respectively. These address conversion operations are performed by `\emph{OFPT\_FLOW\_MOD}' OpenFlow command at the switches to which the nodes are connected. The address conversion overhead at OF switches is negligible. The message overhead for flow-table update between the SDN controller and OF switches depends upon the multiplexing frequency of FRVM mechanism while it can be mitigated by optimizing the frequency of performing multiplexing. 

\section{Conclusion \& Future Work}  \label{sec:conclusion}
From this study, we found the following {\bf key findings}:
\begin{itemize}
\item We proposed a Flexible Random Virtual IP Multiplexing (FRVM) technique as a novel SDN-based MTD mechanism to deal with the network reconnaissance and scanning attacks;
\item We designed and developed a flexible, random IP address mapping technique which is feasible to deploy in SDN environments; 
\item We developed the architecture and communication protocols of FRVM; and
\item We derived probabilistic models to analyze and evaluate the FRVM against random scanning attacks. Our results prove that FRVM can effectively thwart scanning attacks with the attacker success rate close to 1\%. 
\end{itemize}

We plan to conduct the following {\bf future work} items: (1) investigating the effectiveness of FRVM against distributed denial of service (DDoS) attacks; and (2) identifying an optimal multiplexing frequency that can effectively minimizes the overhead derived from maintaining flow-table updates.

\section*{Acknowledgement}
This work was partially supported by US Army Research, Development and Engineering Command (RDECOM) International Technology Center-Pacific (ITC-PAC) and U.S. Army Research  Laboratory (US-ARL)  under  Cooperative  Agreement  FA5209-18-P-0037; The views and conclusions contained in this document are those of the authors and should not be interpreted as representing the official policies, either expressed or implied, of RDECOM ITC-PAC, US-ARL, or the U.S. Government. The U.S. Government is authorized to reproduce and distribute reprints for Government purposes notwithstanding any copyright notation here on.

\bibliographystyle{IEEEtranSN}
\bibliography{mtd_frvm}

\end{document}